\documentclass[
reprint,
superscriptaddress,
preprintnumbers,
amsmath,
amssymb,
prc,
floatfix,
]{revtex4-1}

\usepackage{graphicx}
\usepackage{dcolumn}
\usepackage{bm}
\usepackage[colorlinks,
linkcolor=blue,
anchorcolor=blue,
citecolor=blue,
urlcolor=blue]{hyperref}

\usepackage{amsmath}
\usepackage{xcolor}
\usepackage{soul}
\usepackage{babel}
\usepackage{booktabs}
\usepackage{algorithm}
\usepackage{multirow}
\usepackage{slashed}
\usepackage{graphicx}
\usepackage{dcolumn}
\usepackage{bm}
\usepackage[mathlines]{lineno}

\begin{document}

\title{System-size dependence of jet quenching from light to heavy-ion collisions at LHC}

\author{Man Xie}
\email[]{manxie@wust.edu.cn}
\affiliation{School of Physics and Mechanics, Wuhan University of Science and Technology, Wuhan, Hubei 430065, China}
\affiliation{Key Laboratory of Quark and Lepton Physics (MOE) \& Institute of Particle Physics, Central China Normal University, Wuhan 430079, China}

\author{Xiang-Yu Wu}
\email[]{xiangyu.wu2@mail.mcgill.ca}
\affiliation{Department of Physics, McGill University, Montreal, QC H3A 2T8, Canada}

\author{Han-Zhong Zhang}
\email[]{zhanghz@mail.ccnu.edu.cn}
\affiliation{Key Laboratory of Quark and Lepton Physics (MOE) \& Institute of Particle Physics, Central China Normal University, Wuhan 430079, China}

\author{Xin-Nian Wang}
\email[]{xnwang@ccnu.edu.cn}
\affiliation{Key Laboratory of Quark and Lepton Physics (MOE) \& Institute of Particle Physics, Central China Normal University, Wuhan 430079, China}

\date{\today}

\begin{abstract}
Single inclusive hadron suppression in high-energy heavy-ion collisions provides a sensitive probe of parton energy loss and jet transport coefficient in quark-gluon plasma (QGP). System size dependence of jet quenching can provide further constraints on the dynamics of jet quenching and parton energy loss. Motivated by recent experiments on light-ion collisions at the Large Hadron Collider (LHC), we perform a systematic study of single-inclusive hadron suppression in O+O, Ne+Ne, Xe+Xe and Pb+Pb collisions at the LHC energies. The calculations are carried out within the next-to-leading-order perturbative QCD model, incorporating the nuclear modified initial-state parton distributions, final-state parton energy loss in the higher-twist approach and the temperature dependence of the jet transport coefficient $\hat q/T^3$ from the Bayesian analyses of previous experimental data at the Relativistic Heavy-ion Collider (RHIC) and LHC. Our model calculations can simultaneously describe the measured single-inclusive charged hadron $R_{AA}$ in O+O, Ne+Ne, Xe+Xe and Pb+Pb collisions over a broad range of event centralities and in the minimum-bias events. We further illustrate the system size dependence of jet quenching through the ratio $R_{\mathrm{NeNe}}/R_{\mathrm{OO}}$ and the modification factors $R_{AA}$ as a function of $\langle N_{\mathrm{coll}} \rangle$ which follow a common behavior at fixed $p_{\rm T}$, indicating the formation of QGP even in the smallest collision systems.

\end{abstract}

\maketitle

\noindent {\it \color{blue} Introduction--}
High-energy heavy-ion collisions at the Relativistic Heavy-ion Collider (RHIC) and the Large Hadron Collider (LHC) produce a strongly coupled quark--gluon plasma (QGP) that behaves as an almost perfect fluid \cite{PHENIX:2004vcz,STAR:2005gfr}. Energetic quarks and gluons produced in initial hard scatterings lose energy through interactions with the hot and dense QGP while traversing the medium \cite{Gyulassy:1990ye,Wang:1991xy,Qin:2015srf,Schenke:2009gb,Yazdi:2022bru,Modarresi-Yazdi:2024vfh,Singh:2025duj}. As a result, high-$p_{\rm T}$ hadron production in $A$+$A$ collisions is suppressed relative to the expectation from independent proton--proton ($p$+$p$) collisions, a phenomenon known as jet quenching \cite{Wang:1992qdg}. The magnitude of this suppression is commonly quantified by the nuclear modification factor $R_{AA}$, defined as the ratio of hadron yields in $A$+$A$ and $p$+$p$ collisions scaled by the average number of binary nucleon--nucleon collisions ($\langle N_{\mathrm{coll}}\rangle$) for a given centrality class \cite{Wang:2004yv}. The observed $R_{AA}<1$ for high-$p_{\rm T}$ hadrons and jets at RHIC and the LHC provide the compelling evidence for parton energy loss in the QGP \cite{PHENIX:2008saf,Aamodt:2010jd,Adare:2012wg,CMS:2012aa,Abelev:2012hxa,Aad:2015wga,Khachatryan:2016odn,Acharya:2018qsh}.

In the meantime, $p$+$A$ collisions have been studied extensively to constrain cold nuclear matter (CNM) effects and thereby improve the baseline for isolating the effect of final-state parton energy loss in $A$+$A$ collisions \cite{Salgado:2011wc,Albacete:2016veq,PHENIX:2015fgy,ALICE:2012xs}. Although signatures commonly associated with QGP formation, including a strong elliptic flow \cite{CMS:2010ifv,ALICE:2012eyl,PHENIX:2018lia,ATLAS:2019vcm,STAR:2022pfn} and strangeness enhancement \cite{ALICE:2016fzo}, have been observed in $p$+$A$ collisions, the measured $R_{pA}$ remains close to unity over a broad $p_{\rm T}$ range \cite{ALICE:2012mj,ALICE:2014nqx,ALICE:2021est,CMS:2015ved,ALICE:2015umm,ATLAS:2017pgl,CMS:2016svx}. Moreover, the interpretation of $R_{pA}$ is complicated by uncertainties in determining the average number of binary collisions $\langle N_{\mathrm{coll}}\rangle$ as the normalization factor because of event-selection and geometric biases \cite{Huss:2020dwe,Ke:2022gkq,PHENIX:2023dxl,ALICE:2014xsp}. Similar biases also affect very peripheral $A$+$A$ collisions \cite{ATLAS:2014cpa,ALICE:2018ekf}, complicating the interpretation of the observed centrality dependence of $R_{AA}$. In contrast, minimum-bias (MB) collisions provide a well-defined normalization factor of $A^{2}$ for $R_{AA}$ and serve as an ideal bridge between $p$+$A$ and heavy-ion collisions \cite{Citron:2018lsq,Huss:2020dwe}. The recent CMS measurement of charged-hadron $R_{AA}$ in minimum-bias O+O collisions reports a minimum $R_{AA}\approx0.7$ at $p_{\rm T}\sim6$ GeV, providing evidence for jet quenching in light-ion collisions \cite{CMS:2025bta}. Furthermore, CMS also compared the suppression patterns in O+O, Ne+Ne, Xe+Xe, and Pb+Pb collisions, highlighting the system-size dependence of jet quenching \cite{CMS:2026qef}.

Many theoretical energy-loss-based models have successfully described hadron and jet suppression in heavy-ion collisions, including  Xe+Xe, Au+Au and Pb+Pb collisions \cite{JET:2013cls,JETSCAPE:2022jer,Ehlers:2024miy,Xie:2024xbn}. In contrast, theoretical predictions for hadron and jet suppression in light-ion and $p$+$A$ collisions differ substantially among existing models \cite{CMS:2025bta,CMS:2026qef,Arleo:2020eia,Arleo:2020hat,Huss:2020whe,Xie:2020zdb,Ke:2022gkq,Katz:2019qwv,Liu:2021izt}.  In this work, we perform a systematic study of single inclusive hadron suppression in O+O, Ne+Ne, Xe+Xe and Pb+Pb collisions at the LHC energies within a next-to-leading-order (NLO) perturbative QCD parton model \cite{Owens:1986mp}, incorporating both cold nuclear medium (CNM) effects and final-state parton energy loss with a temperature-dependent jet transport coefficient, $\hat{q}(T)$ from a global analysis of hadron spectra and correlation at both RHIC and LHC using an information field theory-assisted Bayesian inference framework \cite{Xie:2022ght,Xie:2022fak}. 


\noindent {\color{blue}\em NLO parton model with energy loss--}
Within the QCD collinear factorization framework, the cross section for single inclusive hadron production can be expressed as the convolution of parton distribution functions (PDFs), perturbative partonic hard-scattering cross sections, and fragmentation functions (FFs) \cite{CTEQ:1993hwr}. In $A$+$A$ collisions, initial-state nuclear effects are incorporated through nuclear PDFs (nPDFs), while final-state parton energy loss is implemented via medium-modified fragmentation functions (mFFs). The differential yield of single-inclusive hadrons is then given by \cite{Owens:1986mp,Chen:2010te,Liu:2015vna,Zhang:2007ja},
\begin{eqnarray}
	\frac{dN_{AA}}{dyd^2p_{\rm T}} &&=\sum_{abcd} \int d^2r\,dx_a\,dx_b\, t_A(\vec{r})\,t_B(\vec{r}+\vec{b}) \nonumber \\
	&& \hspace{-0.5in}\times f_{a/A}(x_a,\mu^2,\vec{r})\,f_{b/B}(x_b,\mu^2,\vec{r}+\vec{b}) \nonumber \\
	&& \hspace{-0.5in}\times
	\frac{1}{\pi z_c}
	\frac{d\sigma_{ab\rightarrow cd}}{d\hat{t}}
	\tilde{D}_{h/c}(z_c,\mu^2,\Delta E_c)
	+\Delta N_{AA}(\alpha_s^3),
\label{eq:AA-sin-spec}
\end{eqnarray}
where $t_A(\vec{r})=\int dz\,\rho_A(\vec{r},z)$ is the nuclear thickness function with the Woods--Saxon density profile $\rho_A(\vec{r},z)$ \cite{Miller:2007ri}. The impact-parameter dependent nPDFs, $f_{a/A}(x_a,\mu^2,\vec{r})$, are taken from the EPPS16 parameterization \cite{Eskola:2016oht} following Refs.~\cite{Wang:1998ww,Hirano:2003pw,Xie:2022fak}, while the free nucleon PDFs are provided by CT14 parameterization \cite{Hou:2016nqm}. The mFFs $\tilde{D}_{h/c}(z_c,\mu^2,\Delta E_c)$ incorporates parton energy loss, whereas the corresponding vacuum FFs are taken from the Kniehl--Kramer--Potter parameterization \cite{Kniehl:2000fe}. The NLO correction $\Delta N_{AA}(\alpha_s^3)$ at $\mathcal{O}(\alpha_s^3)$ contains $2 \to 3$ processes at the tree level and virtual corrections to the $2 \to 2$ processes.

\begin{figure}
    \centering
    \includegraphics[width=0.9\columnwidth]{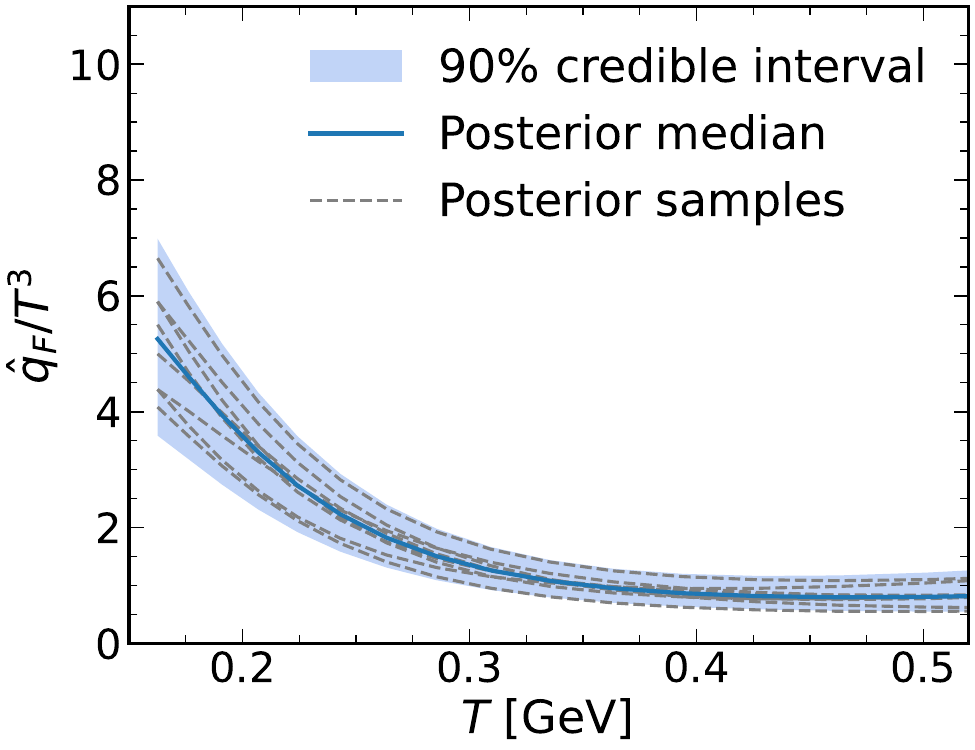}
    \caption{Temperature dependence of $\hat{q}_F/T^3$ obtained from the
global information field theory-assisted Bayesian analysis
\cite{Xie:2022ght,Xie:2022fak}.
The solid blue curve and blue band denote the posterior median
and 90\% credible interval, respectively, while the gray dashed
curves show 8 representative posterior samples.}
    \label{fig_qhat}
\end{figure}

The medium-induced radiative energy loss of a light quark $c$ with four-momentum $p^\mu=(E,\vec{p})$ is computed using the higher-twist formalism \cite{Guo:2000nz,Wang:2001ifa},
\begin{eqnarray}
\frac{\Delta{E}_c}{E} &=& \! \frac{2C_A\alpha_s(E^2)}{\pi} \!\!\int_{\tau_0}^{\infty}\!\!\!\! d\tau \!\!\int \!\! \frac{dl_{\rm T}^2}{l_{\rm T}^2 (l_{\rm T}^2+\mu_D^2)}\!\!\int\!\! dz  \left[1+(1-z)^2\right] \nonumber\\
	&&\times \hat{q}_c(T(\tau)) \frac{p^\mu\cdot u_\mu}{E}
	\sin^2\left[\frac{l_{\rm T}^2(\tau-\tau_0)}{4z(1-z)E}\right],
\label{eq:deltaE}
\end{eqnarray}
where the path integral is evaluated along the jet trajectory starting at $\tau_0=0.6$ fm/$c$. Here, $C_A=3$, $\alpha_s(E^2)$ is the running strong coupling, and $l_{\rm T}$ and $z$ are the transverse momentum and longitudinal momentum fraction of the radiated gluon, respectively. The factor $p^\mu \cdot u_\mu/E$ accounts for the local flow correction, with $u^\mu$ denoting the four-velocity of the QGP fluid. The Debye screening mass is given by
$\mu_D^2=(1+n_f/6)g^2T^2=(1+n_f/6)4\pi\alpha_sT^2$ \cite{Deng:2009ncl,Chang:2014fba,He:2015pra}, where $n_f=3$ is the number of active quark flavors. The jet transport coefficient $\hat{q}_c(T)$ depend on medium temperature $T$ and can provide a good description of hadron data \cite{Xie:2022ght,Xie:2022fak}. The gluon and quark jet transport coefficients are related by the quadratic Casimir factors, $\hat{q}_A=(9/4)\hat{q}_F$. The temperature-dependent jet transport coefficient, $\hat{q}(T)$ has been extracted from a global analysis of single-hadron, dihadron, and $\gamma$--hadron correlation in Au+Au collisions at $\sqrt{s_{\mathrm{NN}}}=200$ GeV and Pb+Pb collisions at $\sqrt{s_{\mathrm{NN}}}=2.76$ and 5.02 TeV using an information field theory-assisted Bayesian inference framework \cite{Xie:2022ght,Xie:2022fak}. Shown in Fig.~\ref{fig_qhat} is the extracted $\hat q_F/T^3$ as a function of the temperature $T$ at the 90\% credible intervals (CI). We use the posterior median of $\hat{q}_F/T^3$ and 8 representative posterior samples in our model calculations here and we neglect the energy dependence of $\hat q_F$.

The space-time evolution of the temperature $T$ and flow velocity for all four collision systems is provided by CLVisc $(3+1)$-dimensional hydrodynamic simulations \cite{Pang:2012he,Pang:2014ipa,Pang:2018zzo,Wu:2021fjf}. 2D TRENTo initial conditions \cite{Moreland:2014oya} for CLVisc hydrodynamic evolution
with the longitudinal envelope parameters for each collision system are tuned to reproduce the measured charged-particle rapidity distributions \cite{ALICE:2018cpu,ALICE:2016fbt}, while the parameters controlling the overall normalization and the shear and bulk viscosity are adjusted to describe the corresponding charged-particle multiplicity, mean transverse momentum, and anisotropic flow.

\begin{figure*}[!tph]
\begin{center}
   \includegraphics[width=1.0\textwidth]{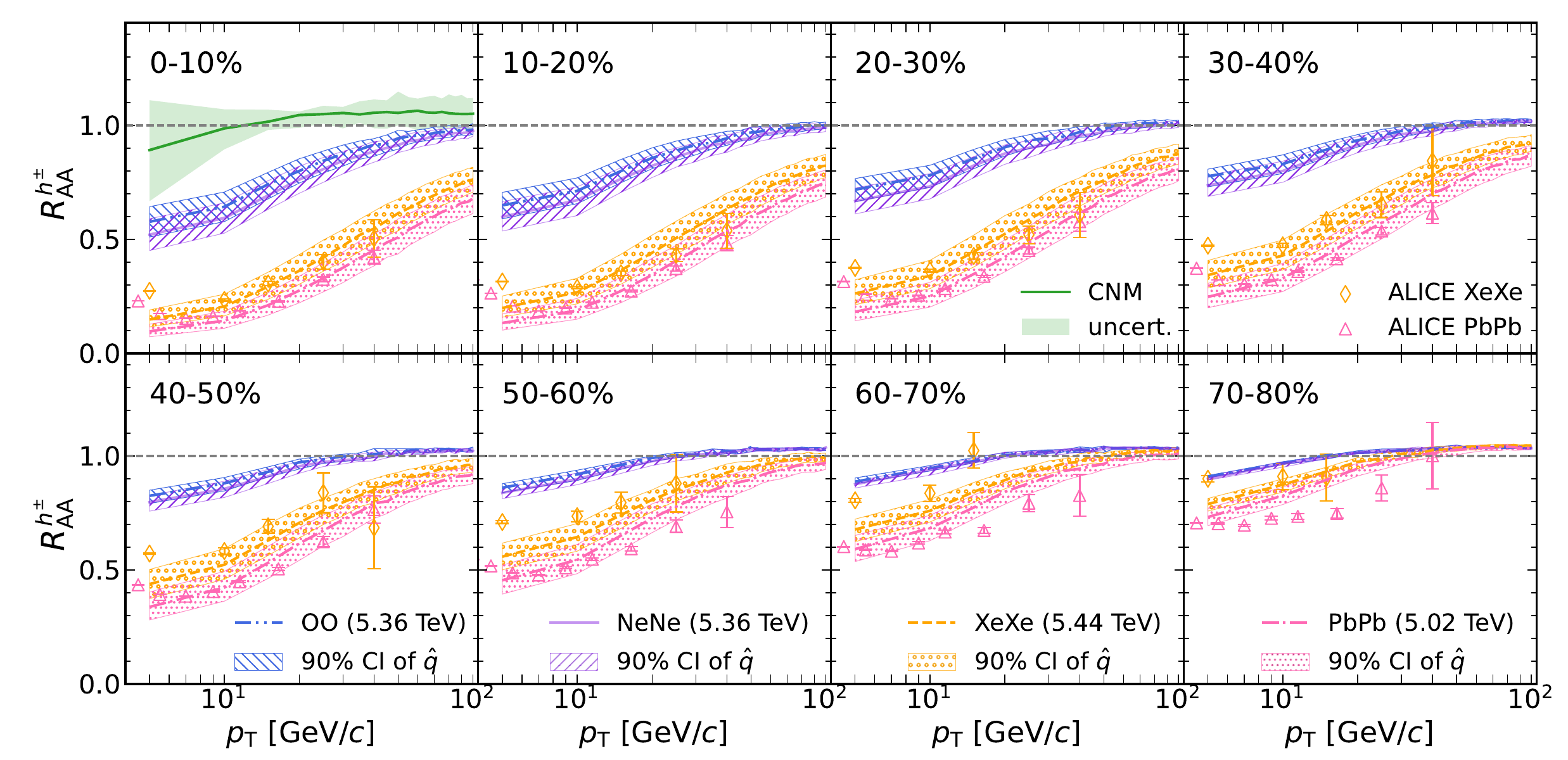}
   \caption{Transverse-momentum dependence of the single-hadron $R_{AA}$ in O+O collisions at $\sqrt{s_{\rm NN}}=5.36$ TeV (blue), Ne+Ne collisions at $\sqrt{s_{\rm NN}}=5.36$ TeV (violet), Xe+Xe collisions at $\sqrt{s_{\rm NN}}=5.44$ TeV (orange), and Pb+Pb collisions at $\sqrt{s_{\rm NN}}=5.02$ TeV (pink) for eight centrality classes from 0--10\% to 70--80\%. Solid curves correspond to the results using the median value of $\hat{q}(T)$, while the shaded bands represent the propagated 90\% credible intervals. Experimental data for Xe+Xe and Pb+Pb collisions \cite{ALICE:2018hza,ALICE:2018vuu} are shown as diamonds and triangles, respectively. In the 0--10\% panel, the green curve and band denote the CNM contribution calculated with the EPPS16 nPDFs for Ne+Ne collisions \cite{Eskola:2016oht}.}
\label{fig:RAA_all}
\end{center}
\end{figure*}

\noindent {\color{blue}\em Nuclear modification factor--}
For a given centrality class, the nuclear modification factor for single inclusive hadron production is defined as \cite{Wang:2004yv}
\begin{eqnarray}
R_{AA} (p_{\rm T}) = \frac{dN_{AA}/dydp_{\rm T}}{T_{AA}(\vec{b})\,d\sigma_{pp}/dydp_{\rm T}} =\frac{dN_{AA}/dydp_{\rm T}}{\langle N_{\rm coll}\rangle\,dN_{pp}/dydp_{\rm T}},
\label{eq:RAA}
\end{eqnarray}
where $T_{AA}(\vec{b})=\int d^2r\,t_A(\vec{r})\,t_B(\vec{r}+\vec{b})$
is the nuclear overlap function at impact parameter $\vec{b}$. The average number of binary collisions is given by
$\langle N_{\rm coll}\rangle=T_{AA}(\vec{b})\,\sigma_{\rm NN}^{\rm in}$,
where $\sigma_{\rm NN}^{\rm in}$ denotes the inelastic nucleon--nucleon cross section \cite{Loizides:2017ack}. 
To calculate the modification factor for minimum-bias events, the full centrality range, $0$--$100\%$, is divided into ten intervals of each width (10\%) using the Glauber model \cite{Glauber:1970jm,Miller:2007ri,Loizides:2017ack}. The minimum-bias $R_{AA}$ is then obtained as the $\langle N_{\rm coll}\rangle$-weighted average,
\begin{eqnarray}
R_{AA}(0\text{--}100\%)=\frac{\sum_i R_{AA,i}\,\cdot \langle N_{\rm coll}\rangle_i}{\sum_i \langle N_{\rm coll}\rangle_i},
\label{eq:RAA_MB}
\end{eqnarray}
where $i=0$--$10\%,\,10$--$20\%,\,\ldots,\,90$--$100\%$.

\noindent {\color{blue}\em Numerical results--}
Using the temperature-dependent $\hat{q}(T)$ together with initial-state nuclear shadowing and final-state parton energy loss, we first calculate medium modification factor $R_{AA}$ of single inclusive hadron spectra as a function of transverse momentum in Xe+Xe (orange) at $\sqrt{s_{\rm NN}}=5.44$ TeV and Pb+Pb (pink) collisions at $\sqrt{s_{\rm NN}}=5.02$ TeV as shown in Fig.~\ref{fig:RAA_all}. The thick lines  correspond to results using the median value of $\hat{q}(T)$ and shaded bands indicate uncertainties at 90\% CI of the $\hat q/T^3(T)$. Results are presented for eight centrality classes (0--10\% to 70--80\%) and compared with the experimental data \cite{ALICE:2018hza,ALICE:2018vuu}. The model provides a good and simultaneous description of the measured $p_{\rm T}$ and centrality dependence of charged-hadron suppression in both collision systems.

 Shown in Fig.~\ref{fig:RAA_all} are also our predicted $R_{AA}$ of single inclusive hadron spectra in O+O (blue) and Ne+Ne (violet) collisions at $\sqrt{s_{\rm NN}}=5.36$ TeV. In the 0--10\% centrality class, the minimum value of $R_{AA}$ is approximately 0.5--0.6 at $p_{\rm T}\sim 5$ GeV in both collision systems. The suppression gradually decreases with increasing transverse momentum, and $R_{AA}$ approaches unity at $p_{\rm T}\sim100$ GeV. To isolate the contribution of jet quenching, we also calculate the corresponding CNM baseline for Ne+Ne collisions, shown in the first panel of Fig.~\ref{fig:RAA_all} as the green curve. The uncertainty is estimated using 1 central and 40 error sets from EPPS16 nPDFs \cite{Eskola:2016oht}. The CNM effects are nearly identical in O+O and Ne+Ne collisions. Shadowing alone suppresses hadron production by approximately 10\% at $p_{\rm T}\sim5$ GeV, whereas anti-shadowing leads to a modest enhancement of about 5\% over the range $20<p_{\rm T}<100$ GeV. In 60--70\% centrality, the CNM effect is smaller. The effects of jet quenching in the predicted $R_{AA}$ in both O+O and Ne+Ne collisions are also very small.

\begin{figure}
    \centering
    \includegraphics[width=0.9\columnwidth]{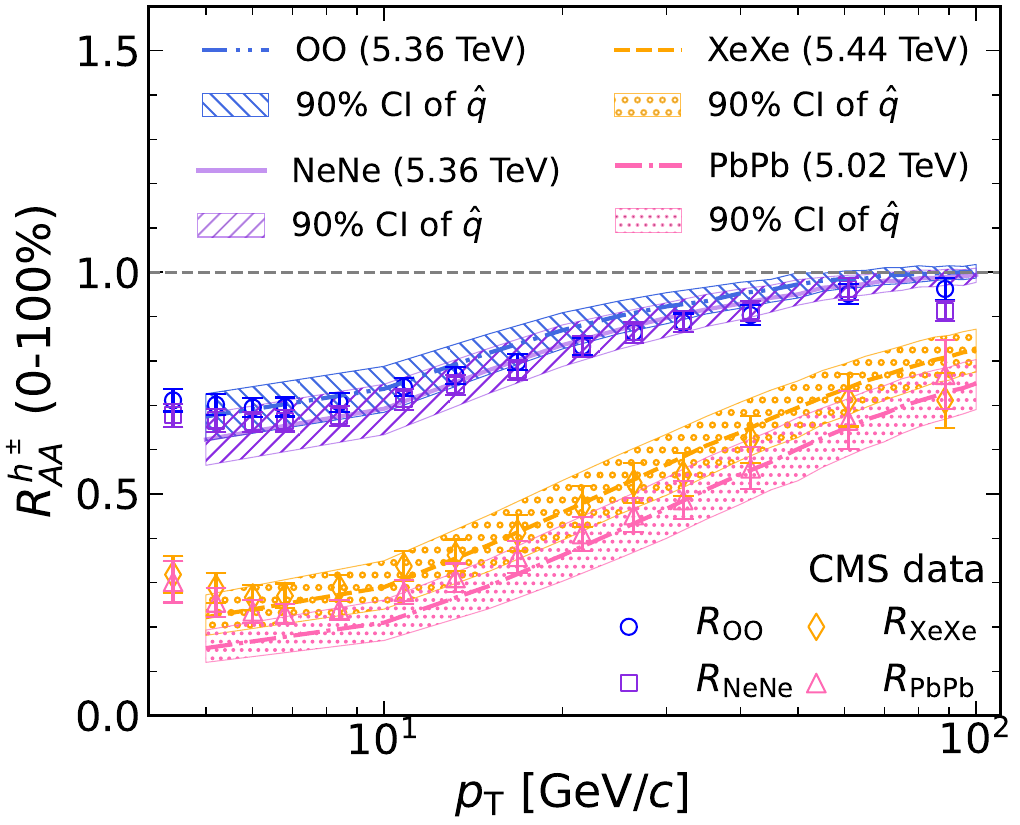}
    \caption{Same as Fig.~\ref{fig:RAA_all}, but for the minimum-bias ($0$--$100\%$) single-hadron $R_{AA}$ compared with the experimental data \cite{CMS:2025bta,CMS:2026qef}. The MB results are obtained as the $\langle N_{\rm coll}\rangle$-weighted average over ten centrality intervals according to Eq.~(\ref{eq:RAA_MB}).}
    \label{fig:RAA_0-100}
\end{figure}

For MB collisions, the normalization factor in the definition of $R_{AA}$ reduces to $A^{2}$ and is therefore free from event-selection and geometric biases \cite{dEnterria:2020dwq,Loizides:2016djv}. Consequently, the MB $R_{AA}$ provides a robust measure of high-$p_{\rm T}$ hadron suppression. We compare our MB numerical results with the available experimental data \cite{CMS:2025bta,CMS:2026qef} in Fig.~\ref{fig:RAA_0-100}.  

Overall, our model calculations are in good agreement with the experimental data. The hadron yields are suppressed by up to approximately 80\% in Xe+Xe and Pb+Pb collisions, compared with about 30\% in O+O and Ne+Ne collisions at $p_{\rm T}\sim 5$ GeV. The calculation slightly overestimates the suppression in Pb+Pb collisions in the $p_{\rm T}\sim5$--10 GeV range, while it slightly underestimates the suppression in O+O collisions at $p_{\rm T}>20$ GeV. This behavior can be traced to the extraction of $\hat{q}(T)$, where the simultaneous inclusion of dihadron observables and low-$T$ constraints favors a somewhat larger jet transport coefficient. When applied to the hotter medium produced in heavy-ion collisions, this results in slightly stronger suppression \cite{Xie:2022ght,Xie:2022fak}. Since the temperature range during the QGP evolution in O+O and Ne+Ne collisions is substantially narrower than that in Xe+Xe and Pb+Pb collisions, the medium suppression in O+O and Ne+Ne is therefore less sensitive to the low-temperature behavior of $\hat{q}(T)$ than in Xe+Xe and Pb+Pb collisions. In any case, the predicted $R_{AA}$ in O+O and Ne+Ne collisions agree well with the experimental data over the whole $p_{\rm T}$ range. The remaining deviations of less than 10\% can be reduced or eliminated by introducing $p_{\rm T}$ dependence of the jet transport coefficient $\hat q_F$.


\begin{figure}
    \centering
    \includegraphics[width=0.9\columnwidth]{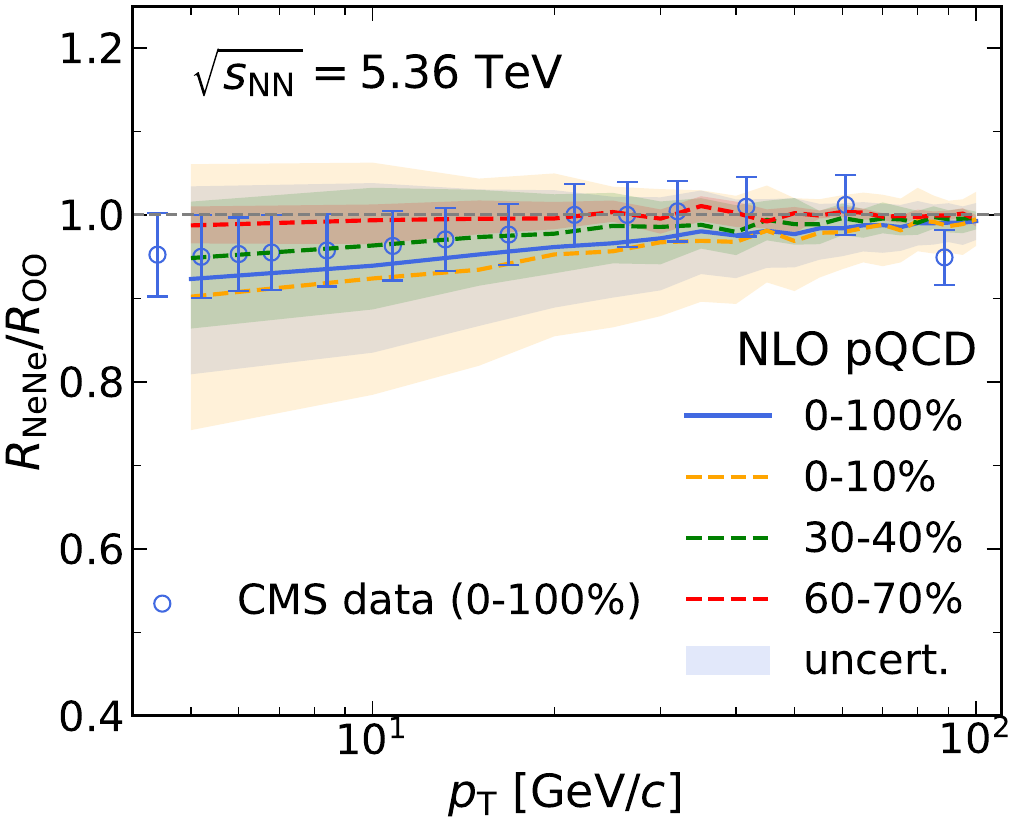}
    \caption{The ratio $R_{\mathrm{NeNe}}/R_{\mathrm{OO}}$ at $\sqrt{s_{\rm NN}} = 5.36 ~\rm TeV$ as a function of $p_{\rm T}$ for the 0--100\%, 0--10\%, 30--40\%, and 60--70\% centrality classes. Solid curves correspond to calculations using the $\hat{q}(T)$-median. The theoretical uncertainty bands and experimental error bars are obtained by propagating the corresponding $R_{AA}$ uncertainties using the standard variance propagation method. The MB results are compared with the experimental data \cite{CMS:2025bta,CMS:2026qef}.}
    \label{fig:RNeO}
\end{figure}

To further quantify the difference in hadron suppression between the two light-ion systems, we calculate the ratio $R_{\mathrm{NeNe}}/R_{\mathrm{OO}}$ for the 0--100\%, 0--10\%, 30--40\%, and 60--70\% centrality classes, as shown in Fig.~\ref{fig:RNeO}. The MB results are compared with the available experimental data \cite{CMS:2025bta,CMS:2026qef}. Our results indicate that hadron suppression in Ne+Ne collisions is approximately 10\% stronger than that in O+O collisions. For all centrality classes, the ratio gradually approaches unity with increasing $p_{\rm T}$, indicating that the difference in jet quenching between the two collision systems becomes progressively smaller at high $p_{\rm T}$.

\begin{figure}
    \centering
    \includegraphics[width=1.05\columnwidth]{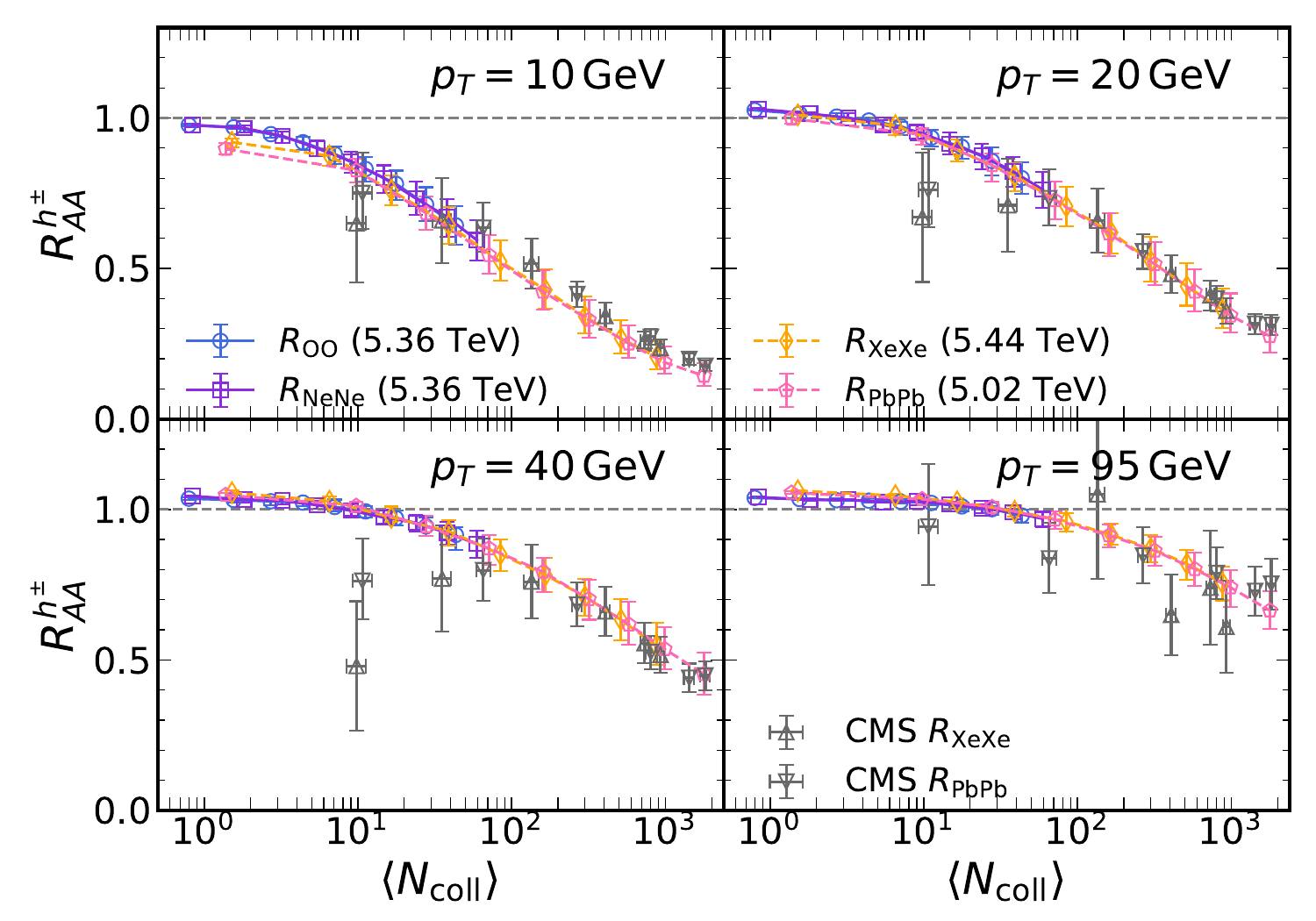}
\caption{The single-hadron $R_{AA}$ as a function of $\langle N_{\rm coll}\rangle$ at fixed hadron transverse momenta $p_{\rm T}=10$, 20, 40, and 95 GeV. Theoretical results for O+O (blue circles), Ne+Ne (purple squares), Xe+Xe (orange diamonds), and Pb+Pb (open hot-pink pentagons) collisions are shown as connected markers. Experimental data for Xe+Xe and Pb+Pb collisions \cite{CMS:2018yyx} are shown as open gray triangles. The theoretical error bars represent the propagated 90\% CI associated with $\hat{q}(T)$.}
\label{fig:RAA_Ncoll}
\end{figure}

Finally, we examine the suppression factor in the four collision systems over a broad range of centralities since they provide an opportunity to investigate the system-size dependence of parton energy loss. We present the $R_{AA}$ as a function of $\langle N_{\rm coll}\rangle$ at several fixed hadron $p_{\rm T}=10$, 20, 40, and 95 GeV, together with the available O+O, Ne+Ne, Xe+Xe and Pb+Pb data \cite{CMS:2018yyx} in Fig.~\ref{fig:RAA_Ncoll}. The calculations reproduce the measured Xe+Xe and Pb+Pb results and exhibit an approximately universal dependence of $R_{AA}$ on $\langle N_{\rm coll}\rangle$ across all four collision systems, especially at large $p_{\rm T}$. For example, the most central O+O and Ne+Ne and 60--70\% Pb+Pb collisions have similar values of $\langle N_{\rm coll}\rangle$. They have nearly identical hadron suppression at fixed $p_{\rm T}$. This universal scaling suggests that $\langle N_{\rm coll}\rangle$ provides an effective measure of the jet-quenching strength at given collision energy and hadron's $p_{\rm T}$ and may serve as a useful indicator of the minimum collision system capable of forming a quark--gluon plasma.

\noindent{\color{blue}\em Summary --}
We have performed a systematic study of single-inclusive hadron suppression in O+O, Ne+Ne, Xe+Xe and Pb+Pb collisions at the LHC energies, within a NLO perturbative QCD model incorporating both initial-state nuclear shadowing and final-state parton energy loss within the higher-twist formalism using a temperature-dependent jet transport coefficient $\hat{q}(T)$.

Our calculations simultaneously reproduce the measured $p_{\rm T}$ and centrality dependence of charged-hadron $R_{AA}$ in Xe+Xe and Pb+Pb collisions and provide a good description of the available O+O and Ne+Ne data. The minimum-bias $R_{AA}$, obtained as the $\langle N_{\rm coll}\rangle$-weighted average over ten centrality intervals, is consistent with the experimental measurements for all four collision systems.

We further quantify the system-size dependence of hadron suppression through the ratio $R_{\rm NeNe}/R_{\rm OO}$ and find that hadron suppression in Ne+Ne collisions is approximately 10\% stronger than that in O+O collisions. More importantly, by studying $R_{AA}$ as a function of $\langle N_{\rm coll}\rangle$ at fixed transverse momentum, we identify an approximately universal scaling behavior across all four collision systems. This observation indicates that $\langle N_{\rm coll}\rangle$ provides a robust proxy of the jet-quenching strength over a broad range of collision systems at similar center-of-mass energies. The observed scaling establishes a simple phenomenological framework for comparing jet quenching from light-ion to heavy-ion collisions and may provide a useful benchmark for identifying the onset of QGP formation in small collision systems.

\begin{acknowledgments}
This work is supported by NSFC under Grant Nos. 12535010, 11935007 and 12605230, the Open Fund of Key Laboratory of Quark and Lepton Physics of Ministry of Education No. QLPL2025P01, and the Faculty Research Startup Fund of Wuhan University of Science and Technology No. 108010001. X.-Y. W. is supported in part by the Natural Sciences and Engineering Research Council of Canada
(NSERC) [SAPIN-2026-00047and SAPIN-2024-00026]. Computations are performed at NSC3.
\end{acknowledgments}

\bibliography{prc_letter}

\end{document}